\documentclass[12pt]{article}
\usepackage{amsmath}
\usepackage{amssymb}
\usepackage{amsthm}
\usepackage[titletoc,title]{appendix}
\usepackage{graphicx}
\usepackage{float}
\usepackage{wrapfig}
\numberwithin{equation}{section}
\usepackage[left=0.92in,right=0.92in,bottom=1.3in]{geometry}
\usepackage{setspace}
\usepackage [linktocpage]{hyperref}
\begin{document}
\title{\bf  Higgs-like (pseudo)Scalars in AdS$_4$, Marginal and Irrelevant Deformations in CFT$_3$ and Instantons on $S^3$\ \\ \ }
\author{{\bf M. Naghdi \footnote{E-Mail: m.naghdi@ilam.ac.ir} } \\
\textit{Department of Physics, Faculty of Basic Sciences}, \\
\textit{University of Ilam, Ilam, West of Iran.}}
\date{\today}
 \setlength{\topmargin}{0.0in}
 \setlength{\textheight}{9.2in}
  \maketitle
  \vspace{-0.1in}
    \thispagestyle{empty}
 \begin{center}
   \textbf{Abstract}
 \end{center}
With a 4-form ansatz of 11-dimensional supergravity over non-dynamical $AdS_4 \times S^7/Z_k$ background, with the internal space as a $S^1$ Hopf fibration on $CP^3$, we get a consistent truncation. The (pseudo)scalars, in the resulting scalar equations in Euclidean AdS$_4$ space, may be viewed as arising from (anti)M-branes wrapping around internal directions in the (Wick-rotated) skew-whiffed M2-branes background (as the resulting theory is for anti-M2-branes) and so, realizing the modes after swapping the three fundamental representations $\textbf{8}_s$, $\textbf{8}_c$, $\textbf{8}_v$ of $SO(8)$. Taking the backreaction on the external and internal spaces, we get massless and massive modes, corresponding to exactly marginal and marginally irrelevant deformations on the boundary CFT$_3$, and write a closed solution for the bulk equation and compute its correction to the background action. Next, considering the Higgs-like (breathing) mode $m^2=18$, having all supersymmetries, parity and scale-invariance broken, by solving the associated bulk equation with mathematical methods, especially the Adomian decomposition method, and analyzing the behavior near the boundary of the solutions, we realize the boundary duals in $SU(4) \times U(1)$-singlet sectors of the ABJM model. Then, introducing new dual deformation $\Delta_+ = 3, 6$ operators made of bi-fundamental scalars, fermions and $U(1)$ gauge fields, we obtain $SO(4)$-invariant solutions as small instantons on a three-sphere with radius at infinity, which actually correspond to collapsing bulk bubbles leading to big-crunch singularities.

\newpage
\setlength{\topmargin}{-0.7in}
\pagenumbering{arabic} 
\setcounter{page}{2} 


\section{Introduction}
Euclidean solutions with finite actions, known as Instantons, as non-perturbative phenomena play important roles in physics from quantum corrections to classical system behavior to early universe cosmology.\footnote{Because our universe may now be or in the past have been in a quasi-stable vacuum state, tunneling from that state to a stable one is interesting. In fact, a false vacuum corresponding to the local minimum of the potential of a scalar field is unstable and decays through tunnelling meditated by instantons or bounces.} In a series of studies- see \cite{Me6}, \cite{Me7}, \cite{Me8}, \cite{Me9}, \cite{Me10} as recent ones- we have presented a number of such solutions in the context of AdS$_4$/CFT$_3$ correspondence, the best model of which is ABJM \cite{ABJM}.

In fact, the ABJM action describes the world-volume of $N$ intersecting M2-branes on an $Z_k$ orbifold of $C^4$ (four complex coordinates), with the orbifold acting as $X^A \to e^{\frac{2 \pi i}{k}} X^A$, $A=1,2,3,4$. In the 't Hooft large $N$ limit and fixed $\lambda= N/k$, the 11-dimensional (11D) supergravity (SUGRA) over $AdS_4 \times S^7/Z_k$ is valid when $N\gg k^5$, and by the orbifold, the subgroup $SU(4) \times U(1) \equiv H$ of the original $SO(8)\equiv G$ remains. The 3D boundary theory is a $U(N)_k \times U(N)_{-k}$ Chern-Simon (CS) gauge theory with $\mathcal{N}=6$ supersymmetry (SUSY) and matter fields (scalars $Y^A$ and fermions $\psi_A$) in bi-fundamental representations (reps) ($\textbf{4}_1$ and $\bar{\textbf{4}}_{-1}$) of $H$.

Here, by keeping the ABJM background geometry unchanged and taking a 4-form ansatz of the 11D SUGRA, composed of the ABJM internal $CP^3 \ltimes S^1/Z_k$ space elements and scalars in the external 4D Euclidean anti-de Sitter ($EAdS_4$) space, associated with probe (anti)M-branes wrapped around mixed directions in (M2-branes)anti-M2-branes background resulting in anti-M2-branes theory, we will have a \emph{consistent truncation} in that just $H$-singlet fields remain in the truncated theory and all dependencies on the internal 7D space are omitted in resulting equations.\footnote{It is discussed in \cite{Duff1985-2} that a consistent truncation includes the singlet fields under the internal isometry group, by setting to zero non-singlets ones; see also \cite{Bremer1998}. Particularly, a consistent truncation of M-theory over $AdS_4 \times S^7$ to the 4D $\mathcal{N}=8$ $SO(8)$ gauged supergravity is presented in \cite{Duff99} including a special case, where just the graviton and a scalar potential is retained, as is the case here; For a newer look, see \cite{Gauntlett03}.}

On the other hand, instantons as topological objects should not backreact on the background geometry. To this end, we solve the truncated equations in $EAdS_4$ together with the equations resulting from zeroing the energy-momentum (EM) tensors of the Einstein's equations that result in equations for massless and massive bulk (pseudo)scalars, which in turn correspond to exact and irrelevant marginal deformation of the dual boundary theory. In addition, we consider a Higgs-like mode ($m^2=18$), already known as \emph{breathing mode}; see \cite{Bremer1998}, \cite{Hawking1998}, \cite{Bak-Yun2010} and \cite{Gauntlett03}. To get solutions for the bulk scalar equations, we employ the usual mathematical methods and especially the Adomian Decomposition Method (ADM) \cite{Adomian1994} to solve the Nonlinear Partial Differential Equations (NPDEs). 

Next, after analyzing the bulk solutions near the boundary and dual symmetries, we propose the corresponding dual operators to deform the boundary action with and find solutions. In fact, because the bulk setups and solutions break all SUSYs, $\mathcal{N}=8 \rightarrow 0$, parity- and scale-invariance, to realize the boundary duals, we swap the three fundamental reps $\textbf{8}_s$, $\textbf{8}_c$ and $\textbf{8}_v$ of $SO(8)$. With such a swapping, we could realize the $H$-singlet scalars and pseudoscalars in the mass spectrum of the 11D SUGRA on the background geometry after the branching $G \rightarrow H$, corresponding to $H$-singlet boundary operators. Meanwhile, as the scale symmetry is violated because of the mass term in the equations and their extreme nonlinearity, the solutions should preserve the $SO(4)$ symmetry of the original isometry $SO(4,1)$ of $EAdS_4$. Keeping a singlet sector of the boundary ABJM action,\footnote{As discussed in \cite{Me9}, the boundary solutions might also realize in singlet sectors of 3D $U(N)$ and $O(N)$ CS matter theories.} with only one scalar, one fermion and $U(1) \times U(1)$ part of the gauge group, we find such an $SO(4)$-invariant solutions with finite actions, which are always small instantons triggering instabilities on a three-sphere with radius $r$ at infinity ($S_\infty^3$), and describe big crunch singularities in the bulk. 

This article is organized as follows: In section 2, we present the 11D SUGRA background, including the 4-form ansatz, and the equations for (pseudo)scalars in $EAdS_4$; Emphasizing that in the anti-M2-branes background, a (pseudo)scalar becomes Higgs-like providing spontaneous symmetry breaking and making the main equation homogeneous. In subsection 2.1, we consider backreaction; that is after computing the EM tensors of the Einstein’s equations in Appendix A, from zeroing them and solving the resulting scalar equations with the main one in the bulk, we get solvable PDEs for the massless and massive (pseudo)scalars from taking the backreaction on the external and internal spaces, respectively.  Next, in subsection   
2.2, we present an exact solution for the equations in the latter subsection and compute its corrections to the background action. In section 3, we employ known methods of differential equations to solve the main Higgs-like NPDE and get solutions near the boundary. In particular, in subsection 3.1, we use the ADM (with details in Appendix B) to obtain solutions appropriate for near the boundary analyzes of the Higgs-like mode $m^2=18$, up to the third order of the perturbative series expansion. In section 4, we first discuss dual symmetries from the bulk setups, equations and solutions and next, deal briefly with the spectrum of 11D SUGRA over $AdS_4 \times S^7/Z_k$ and check whether we can find the desired $H$-singlet scalars and pseudoscalars among various generations after swapping the fundamental reps of $SO(8)$ for gravitino and the branching of $G \rightarrow H$; Then, we present basic elements of AdS$_4$/CFT$_3$ correspondence for (pseudo)scalars needed for our boundary analyzes. In section 5, we look for dual solutions in ABJM-like 3D field theories; and in this way, in subsections 5.1 and 5.2,  we consider marginal and irrelevant deformations with new $H$-singlet $\Delta_+ =3, 6$ operators, corresponding to the massless (when taking the backreaction) and massive bulk states, respectively, and find $SO(4)$-invariant solutions with finite actions as instantons. In addition, from the boundary solutions, we confirm the state-operator correspondence, match the bulk-boundary parameters and determine an unknown scalar function in a bulk solution from the correspondence. Meanwhile, we remind that with a marginal triple-trace deformation of a dimension-one operator composed of bi-fundamental scalars, we could build the tri-critical $O(N)$ model and find Fubini-like instantons. Also, we will confirm the Bose-Fermi (BF) duality between a deformation with the latter $\Delta_+ =3$ operator (in fact the massless Regular Boson (RB) model) and a deformation with a $\Delta_+ =6$ operator composed of bi-fundamental fermions (in fact the massless Critical Fermion (CF) model) at least at the level of solutions and correspondence, in subsection 5.2. Eventually, in section 6, we present a summary along with comments on solutions, physical interpretations, connections with other studies and related issues.

\section{From 11D Supergravity to 4D Gravity Equations}\label{section02}
We start with the 4-form ansatz \footnote{see also \cite{Me6}.}
\begin{equation}\label{eq01}
  G_4 = f_1\, G_4^{(0)} + R^4\, df_2 \wedge J \wedge e_7 + R^4\, {f}_3\, J^2 
\end{equation} 
for 11D SUGRA over $AdS_4 \times S^7/Z_k$ when the internal space is considered as a $U(1)$ bundle on $CP^3$, where $ G_4^{(0)}=d{\mathcal{A}}_3^{(0)} = N \mathcal{E}_4$ is for the ABJM \cite{ABJM} background with $N=(3/8) R^3$ units of flux quanta on the internal space, $R=2 R_{AdS}$ is the $AdS$ curvature radius, $\mathcal{E}_4$ is the unit-volume form on $AdS_4$, $J$ is the K\"{a}hler form on $CP^3$, $e_7$ is the seventh vielbein \footnote{It is noted that the vielbein is along the fiber direction when we view $S^7/Z_k$
as $S^1$ fibration over $CP^3$.} (of the internal space) and $f_i$'s with $i=1,2,3$ are scalar functions in bulk coordinates.

Taking the ansatz (\ref{eq01}), from the Bianchi identity and Euclidean 11D equation 
\begin{equation}\label{eq02}
  dG_4=0, \qquad d \ast_{11} G_4- \frac{i}{2}\, G_4 \wedge G_4=0, 
\end{equation}
we get
\begin{equation}\label{eq03}
   f_3= f_2 \pm \frac{C_2}{R}, \qquad \bar{f}_1 = i\, \frac{3}{16}\, R^5\, f_3^2\, \pm\, i\, \frac{3}{8} C_3\, R^3,
\end{equation} 
where ${C}_i$'s are real constants, $N\, f_1 \equiv \bar{f}_1$ and note that the plus and minus signs on the last term of the RHS equation indicate considering the Wick-rotated (WR) and skew-whiffed (SW) backgrounds respectively, and that the ABJM background realizes with $C_3=1$. In addition, from the equation (\ref{eq02}), making use of (\ref{eq03}), we get
\begin{equation}\label{eq04}
 \Box_4 f_3 - \frac{1}{R_{AdS}^2} \left(1 \pm 3\, {C}_3 \right) {f}_3 - 6\, {f}_3^3=0,
\end{equation}
\footnote{Note that the so-called $\phi^4$ coupling constant here is $\lambda_4=3$ (given that $\lambda = 2  \lambda_4$); see for instance \cite{Petkou2003} also for discussions on the zeroing of the scalar third-order self-interaction.} where $\ast_4 d \left(\ast_4 d{f}_3 \right)=\Box_4$ is the $EAdS_4$ Laplacian, and we use the following conventions:
\begin{equation}\label{eq05}
\ast_4 \textbf{1} =\frac{R^4}{16} \mathcal{E}_4, \quad \ast_7 \textbf{1} =\frac{R^7}{3!} J^3 \wedge e_7=R^7\, \mathcal{E}_7, \quad \ast_7 \left(J \wedge e_7 \right)= \frac{R}{2} J^2.
\end{equation} 

Next, from (\ref{eq03}) and (\ref{eq04}), we write \footnote{See \cite{Me9}, \cite{Me10} and \cite{Me11} for similar equations.} 
\begin{equation}\label{eq08}
     \Box_4 f_2 - m^2\, f_2 \mp \delta\, f_2^2 - \lambda\, f_2^3 \mp F=0,
\end{equation}
\footnote{It should be noted that the plus sign in front of the terms containing $C_2$ shows that the true vacuum is placed on the right-hand side (RHS) of the false vacuum in the corresponding double-well potential, and vice versa for the minus sign.} where 
\begin{equation}\label{eq08a}
m^2=\frac{4}{R^2} \left(1 \pm 3\, C_3 + \frac{9}{2}\, C_2^2 \right), \quad \delta = \frac{18}{R}\, C_2,  \quad \lambda=6, \quad F = \frac{4}{R^3}\left(C_2 \pm 3\, C_2\, C_3 + \frac{3}{2}\, C_2^3 \right).
\end{equation}  

 To make the equation (\ref{eq08}) homogeneous (that is $F=0$), besides $C_2=0$, we have to set
\begin{equation}\label{eq09}
\delta^2= 27\, m^2 \ \ \  \texttt{OR} \ \ \ C_2^2 = \frac{\mp 6\, C_3 -2}{3} 
\end{equation}
and so 
\begin{equation}\label{eq09a}
m^2 R_{AdS}^2= -2 (1 \pm 3\, C_3) \equiv -2\, \bar{m}^2 R_{AdS}^2,
\end{equation}
where $\bar{m}^2$ is indeed the squared mass of $f_3$ in (\ref{eq04}). To have physically permissible (non-imaginary) masses in this case, we have just to consider the SW version with $C_3 \geq 1/3$; and as a result, the SW ABJM background realizes with $C_3=1$ and then $m^2 R_{AdS}^2= +4$ ($C_2 =2/\sqrt{3}$).

In addition, as noticed in \cite{Me10}, $\pm \left(C_2/2 \right) =\pm \sqrt{{-\bar{m}^2}/{\lambda}}$ are in fact homogenous vacua and so, the (pseudo)scalar is Higgs-like and the LHS relation in (\ref{eq03}) imposes spontaneous symmetry breaking, where $f$ acts as fluctuation around the homogeneous vacua.

\subsection{Taking Backreaction and Resulting Equations} \label{section02.01}
To take backreaction, we should first compute the EM tensors of the corresponding Einstein's equations, the details of which are in Appendix \ref{Appendix.A1}. In fact, since we are looking for instantons that, as topological objects, should not backreact on the background geometry, we solve the main bulk equations with the equations in Appendix \ref{Appendix.A1}, resulting from zeroing the EM tensors. 

In this way, we first see that the equation (\ref{eqA13}) is solved with (\ref{eq08}) with  
\begin{equation}\label{eq10}
\Box_4\, f_{2}=0,
\end{equation} 
which means taking the backreaction of the external $AdS_4$ space on the background geometry gives the \emph{massless} $m^2 R_{AdS}^2=0$ bulk (pseudo)scalar replying to the boundary \emph{exactly marginal} operators.\footnote{See \cite{Me8}, \cite{Me9}, \cite{Me3} and \cite{Me4} for discussions on such a correspondence.}

As the same way, noting that the equation (\ref{eqA14a}) is the same as the main one (\ref{eq08}), from solving the equations (\ref{eqA14}) as well as (\ref{eqA13}) and (\ref{eqA14}) with (\ref{eq08}), that is taking the backreaction of the internal (indeed $CP^3$) and whole 11D space, we get
\begin{equation}\label{eq17}
\Box_4 f_2 - \frac{2}{R^2}\, f_2 = \pm\, \frac{2\, C_2}{R^3},
\end{equation}
\begin{equation}\label{eq18}
\Box_4 f_2 - \frac{8}{9\, R^2}\, f_2 = \pm\, \frac{8\, C_2}{9\,R^3},
\end{equation} 
respectively, with $m^2 R_{AdS}^2=1/2, 2/9$ corresponding to the \emph{marginally irrelevant} $\Delta_{\pm} = {3}/{2} \pm {\sqrt{11}}/{2}, \, {3}/{2} \pm {\sqrt{(89/9)}}/{2}$ boundary operators, which we encountered the former recently in \cite{Me9}. 

\subsection{A Solution For the Case with Backreaction} \label{section02.02}
One may solve the equations (\ref{eq17}) and (\ref{eq18}) using the usual mathematical methods, such as separation in variable. However, a well-known closed solution for the equations- leaving out the inhomogeneous terms that do not contribute to the dynamics \footnote{Note also that to adjust (\ref{eq17}) and (\ref{eq18}) with the main equation (\ref{eq08}) with (\ref{eq08a}), we have to set $C_2=0$.}- reads \cite{Witten}, \cite{Freedman1998}
\begin{equation}\label{eq22}
     f_0(u,\vec{u})= \bar{C}_{\Delta_+} \left(\frac{u}{u^2+(\vec{u}-\vec{u}_0)^2} \right)^{\Delta_+}, \quad \bar{C}_{\Delta_+}=\frac{\Gamma(\Delta_+)}{\pi^{3/2}\, \Gamma(\nu)}.
\end{equation}
where $\Delta_+$ ($\Delta_-$) is the larger (smaller) root  of $m^2=\Delta (\Delta-3)$ in $AdS_4$, with $\Delta_{\pm}=3/2 \pm \nu$, $\sqrt{9 + 4\, m^2}=2 \nu$, and we use the $EAdS_4$ metric 
\begin{equation}\label{eq20}
 ds^2_{EAdS_4} = \frac{R^2}{4\, u^2} \left(du^2 + dx^2 + dy^2 + dz^2 \right),
\end{equation}
noting $\vec{u}=(x,y,z)$, in upper-half Poincar$\acute{e}$ coordinates and so
\begin{equation}\label{eq21}
     \Box_4 f = \frac{4\, u^2}{R^2} \left(\partial_i \partial_i + \partial_u \partial_u - \frac{2}{u} \partial_u \right) f.
\end{equation} 

On the other hand, because the bulk solutions including the backreaction correspond to variants of marginal operators, for simplicity we consider the instanton solution for (\ref{eq10}) and compute its correction to the background action. To this end, as the background geometry does not change, we use the right parts of the bosonic action of 11D SUGRA in Euclidean space as
\begin{equation}\label{eq23} 
  S_{11}^E = -\frac{1}{4 \kappa_{11}^2} \int \left({G}_4 \wedge \ast_{11} {G}_4 - \frac{i}{3}\, {\mathcal{A}}_3 \wedge {G}_4 \wedge {G}_4 \right),
\end{equation}
where $\kappa_{11}^2 = 9 \pi\, \mathcal{G}_{11}=\frac{1}{4\pi} (2\pi l_p)^9$, and $\kappa_{11}$, $\mathcal{G}_{11}$ and $l_p$ are the 11D gravitational constant, Newton's constant and Plank length, respectively.

Next, from the ansatz (\ref{eq01}), using (\ref{eq05}), we write its 11D dual 7-form as 
\begin{equation}\label{eq24}
  G_7 = \frac{8}{3} R^3\, \bar{f}_1\, J^3 \wedge e_7 - \frac{R^5}{2} \ast_4 df_2 \wedge J^2 + \frac{R^7}{8} f_3\, \mathcal{E}_4 \wedge J \wedge e_7;
\end{equation}
and 
\begin{equation}\label{eq25}
 {G}_4 =  d{\mathcal{A}}_3, \quad {\mathcal{A}}_3 = \tilde{\mathcal{A}}_3^{(0)} + R^4\, \left(f_3\, J \wedge e_7 \right), \quad \tilde{G}_4^{(0)} = d\tilde{\mathcal{A}}_3^{(0)} = \bar{f}_1\, \mathcal{E}_4.
\end{equation} 
By placing the latter relations in (\ref{eq23}), using (\ref{eq03}) (noting that with $C_2=0$, $f_2=f_3$), we get 
\begin{equation}\label{eq26}
  \tilde{S}_{11}^E = - \frac{R^{9}}{16\, \kappa_{11}^2} \int \bigg[-\frac{3}{2}C_3^2\, \mathcal{E}_4 + df_3 \wedge \ast_4 df_3 + \frac{R^2}{2} f_3^2\, \mathcal{E}_4 + \frac{3}{8} R^4 f_3^4\, \mathcal{E}_4 +i \frac{4}{3 R} d\left(f_3^2\, \mathcal{A}_3^{(0)} \right) \bigg] \wedge J^3 \wedge e_7,
\end{equation}
where the first term on the RHS is the contribution of the ABJM background realized with $C_3=1$, as one may see from the second term on the RHS relation in (\ref{eq03}), and the last (surface) term, as a total derivative, does not contribute to the equations and we discard it. 

Then, to compute the action (\ref{eq26}) based on the solution (\ref{eq22}) with $\Delta_+ =3$, we use 
\begin{equation}\label{eq27}
\mathcal{E}_4 = -\frac{du}{u^4} \wedge  dx \wedge dy \wedge dz, \quad \texttt{vol}_7 = \frac{R^7}{3!} \int J^3 \wedge e_7= \frac{\pi^4\, R^7}{3\, k}, \quad \kappa_{11}^2 = \frac{16}{3}  \left(\frac{\pi^{10}\, R^9}{3\, k^3} \right)^{1/2},
\end{equation}
and the 3D spherical coordinates, setting $|\vec{u}-\vec{u}_0|=r$. As a result, the finite contribution of the action, after integrating on the external space coordinates, in the unit 7D internal volume, reads 
\begin{equation}\label{eq28}
  \bar{S}_{11}^{corr.} = - \hat{c}\, \sqrt{\frac{k^3}{R}} \frac{1}{\epsilon^6} \left(1 + \check{c}\, \frac{R^2}{\epsilon^{6}} \right),
\end{equation} 
where $\hat{c}\simeq 0.000016$ and $\check{c}\simeq 0.0033$, and because of singularities, we have included $\epsilon>0$ as a cutoff parameter to evade the infinity of integrals with respect to (wrt) $u$; see \cite{Bianchi3}. Meantime, we note that for finite $k$ and $R$, (\ref{eq28}) is a small contribution. 

\section{Solutions For the Higgs-Like Scalar Equation} \label{section03}
The Higgs-like (pseudo)scalar equation of (\ref{eq08}), with (\ref{eq09}) and (\ref{eq21}), reads \footnote{From now on, we use $f_2 \equiv f$ and the plus sign for the $f^2$ term in the equations.}
\begin{equation}\label{eq29}
 \left[\partial_i \partial_i + \partial_u \partial_u - \frac{2}{u} \partial_u - \frac{m^2}{u^2} \right] f(u,\vec{u})+ \frac{1}{u^2} \left[3 \sqrt{3}\, m\, f(u,\vec{u})^2 - 6\, f(u,\vec{u})^3 \right]=0.
\end{equation}
For its linear part, using the spherical coordinates with $r=|\vec{u}|$, discarding the angular parts, and separation of variables, $f_0(u,r)=f(r) g(u)$, we can write 
\begin{equation}\label{eq30}
    \left[\frac{d^2}{d r^2} + \frac{2}{r} \frac{d}{d r} - k^2 \right] f(r)=0, \quad \left[\frac{d^2}{d u^2} - \frac{2}{u} \frac{d}{d u} - \frac{m^2}{u^2}+ k^2 \right] g(u)=0;
\end{equation}   
with combinations of Hyperbolic and Bessel (or with $k=i \kappa$, Trigonometric and Modified Bessel) functions as solutions for $f(r)$ and $g(u)$, respectively.\footnote{An interesting solution for the $r$ part is $f(r) \sim e^{-r}/r$, which might be considered as constrained instantons; see for instance \cite{KubyshinTinyakov01} and also \cite{Me9}, where we discussed a similar solution in a 3D boson model. Another interesting solution is
\begin{equation}\label{eq31}
    f(r)= \tilde{C}_1\, r^\ell + \frac{\tilde{C}_2}{r^{\ell+1}},
\end{equation}
where $\tilde{C}_1$ and $\tilde{C}_2$ are constants and $\ell (\ell+1)=\kappa^2$.} Then, one may use the leading order (LO) solutions to get the higher-order solutions of the full NPDE. The resulting solutions always reproduce the right behavior of (pseudo)scalars near the boundary as
\begin{equation}\label{eq32}
f(u\rightarrow 0, r) \approx \alpha(r)\, u^{\Delta_-} + \beta(r)\, u^{\Delta_+}.
\end{equation}

On the other hand, one may employ an ansatz like 
\begin{equation}\label{eq33}
   f(u,r)=F(\xi), \quad \xi=u^{1/2}\ f(r),
\end{equation}
which turns (\ref{eq29}) into the following NODE
\begin{equation}\label{eq34}
 \left[ \frac{d^2}{d \xi^{2}}-\frac{5}{\xi}\, \frac{d}{d\xi} -\frac{4\, m^2}{{\xi}^{2}} \right] F(\xi)-\frac{4}{{\xi}^{2}}\, \mathcal{F}(F(\xi)) =0,
\end{equation}
where we define
\begin{equation}\label{eq35}
    \mathcal{F}(F(\xi)) \equiv - 3 \sqrt{3}\, m\, F(\xi)^{2} + 6\, F(\xi)^{3}.
\end{equation}
As a result, the appropriate part of a perturbative solution for (\ref{eq34}), up to the first or next-to-leading order (NLO), reads \footnote{It is recalled that we generally use
\begin{equation}\label{eq36}
f^{(n)} = \sum_{n=0}^n f_n.
\end{equation}}
\begin{equation}\label{eq37}
    f^{(1)}(u,r) = \sum_{{l}=-}^+ \, C_{{l}}\, \left( u\, f(r)^2 \right)^{\Delta_{{l}}}, 
\end{equation}
where $C_{{l}}$'s are real constants.

As the same way, by $\xi=r/u$ (the so-called self-similar reduction method via the scale-invariance of the variables; see for instance \cite{Polyanin2012}), the equation (\ref{eq29}) turns into
\begin{equation}\label{eq38}
 \left[ \left({\xi}^{2}+1 \right) {\frac{{d }^{2}}{{d}{\xi}^{2}}} + \frac{(2+ 4\, \xi^2)}{\xi}\,\frac{d}{d \xi}- m^2 \right] {F}(\xi) - \mathcal{F}(F(\xi)) =0.
\end{equation}
A solution for the linear part of the latter equation is in terms of Legendre functions, and from that, one may build perturbative series solutions up to higher-orders; for such a solution, see \cite{Me9}. Alternatively, we can also use 
\begin{equation}\label{eq39}
 F(\xi)= e^{\int\, G(\xi)\, d\xi}, \quad \frac{1}{F(\xi)}\frac{dF(\xi)}{d\xi}=G(\xi), 
\end{equation}
which turns the equation (\ref{eq38}) into the following first-order \emph{Riccati equation}
\begin{equation}\label{eq40}
\left({\xi}^{2}+1 \right) \left[ \frac{dG(\xi)}{d\xi} + G(\xi)^2 \right]+\frac{1}{{\xi}} (2+ 4\, \xi^2)\, G(\xi) - m^2  =0.
\end{equation}
For massive modes, a common series solution for the latter equation, keeping the normalizable term appropriate for the corresponding boundary analyzes of AdS$_4$/CFT$_3$, reads
\begin{equation}\label{eq41}
 f_0(u,r) = \tilde{C}_{\Delta_+} \left(\frac{u}{r} \right)^{\Delta_+},
\end{equation}
from which one may build higher-order solutions. For example, for the mode $m^2=18$ that we consider, a series expansion around $u=0$, up to NLO, reads 
\begin{equation}\label{eq42}
  f^{(1)}(u,r) = \left[ \check{C}_{\Delta_-}\, \ln(\frac{r}{u}) \right] \left(\frac{u}{r}\right)^{\Delta_-=-3}
         + \hat{C}_{\Delta_+} \left(\frac{u}{r}\right)^{\Delta_+=6}
\end{equation}
with the real constants $\check{C}_{\Delta_-}$ and $\hat{C}_{\Delta_+}$- when doing boundary analyzes, we return to this solution as well.

\subsection{Solutions of the Equation for \pmb{$m^2=18$} with ADM} \label{sub03-01}
Here we employ the ADM formulation in Appendix \ref{Appendix.A2}, to build series solutions appropriate for near the boundary analyzes of the specific Higgs-like mode $m^2=18$. This mode could be realized with $C_3=17/3$ in the WR version of (\ref{eq04}) (equally for (\ref{eq08}) in addition to $C_2=0$) and with $C_3=10/3$, $C_2=\sqrt{6}$ in the SW version of \ref{eq08}) with $F=0$. As a result, the series solutions of these equations about $u=0$, with the initial or near the boundary  data from (\ref{eqA15}) with $\Delta_+=6$ and the Adomian polynomials (\ref{eqA18}) with $\delta=0$ and $\delta= 9\, \sqrt{6}$,  respectively, up to NNNLO, wrt (\ref{eq36}), read
\begin{equation}\label{eq43} 
   f^{(3)}(u,r)= -5 f(r) \left[1+20\,\ln(u) + 400\, \ln(u)^2-8000\, \ln(u)^3  \right] {u}^{6} +  O({u}^{8}),
\end{equation}
\begin{equation}\label{eq44}
   f^{(3)}(u,r)= -20\,f(r)\, {u}^{6} + \frac{17965}{43904} \left(\frac{d^2 f(r)}{d r^2} + \frac{2}{r} \frac{d f(r)}{d r} \right) {u}^{8} + O({u}^{10}).
\end{equation}
Meanwhile, from near the boundary behavior of the closed solution of (\ref{eq22}),
\begin{equation}\label{eq45}
    f_0(u\rightarrow0,r) \approx \bar{C}_{\Delta_+}\, \left( \frac{u}{r^{2}} \right)^{\Delta_+},
\end{equation}
we can read $f(r)=\bar{C}_6/{r^{12}}$ to rewrite the series solutions clearly.

On the other hand, we can use (\ref{eqA21}) with (\ref{eqA21a}) and near the boundary behavior of the closed solution of (\ref{eqA20}),
\begin{equation}\label{eq46}
g_0(u\rightarrow 0,r) = \frac{2}{\sqrt{3}} \frac{{b}_0}{(a_0^2-b_0^2+ r^2)} \left[1 - \frac{2\, a_0}{(a_0^2-b_0^2 + r^2)}\, u \right], 
\end{equation}
as the initial data, which might also be read from the LHS relation in (\ref{eqA15}), in the ADM, to get approximate solutions. As a result, we arrive at a series solution about $u=0$, up to the first iteration of ADM or NLO of the expansion, as
\begin{equation}\label{eq47}
f^{(1)}(u,r) = \sum_{\Delta_+=1}^6\,  \frac{\mathcal{H}_{\Delta_+}(r, a_0, b_0, m)\, u^{\Delta_+}}{(a_0^2-b_0^2 + r^2)^{\Delta_+}},
\end{equation}
where $\mathcal{H}_{\Delta_+}(r, a_0, b_0, m)$ is a polynomial of its arguments; and in particular, for the term corresponding to the bulk mode $m^2=18$, it becomes
\begin{equation}\label{eq47a}
\mathcal{H}_{6} = - \frac{64}{\sqrt{3}}\, a_0^3\, b_0^3.
\end{equation}

\section{Dual Symmetries, Mass Spectrum and Correspondence} \label{section04}
First, we remind that the truncation here is consistent, considering that our ansatz (\ref{eq01}) is $H$-singlet, given that $e_7$, $J$ and the (pseudo)scalars in resulting equations respect the same symmetry. Second, the setups here are as if we add $\ell$ probe (anti)M-branes to the (WR)SW M2-branes background and so, the resultant theory is for anti-M2-branes with the quiver gauge group of $SU(N+\ell)_k \times SU(N)_{-k}$. Indeed the (anti)M-branes wrap around mixed internal and external directions and so break all SUSY's and parity \footnote{According to \cite{DuffNilssonPope84}, when all components of the 11D 4-form are in the internal space, such a thing happens as well; and according to \cite{Page1984}, with such $G_4$, the resulting solutions are unstable.}, and that to realize the latter we focus on $U(1) \subset U(\ell)$ part of the gauge group (in the large $k$ limit) and keep $G$ as a spectator- a so-called novel Higgs mechanism; see for instance \cite{ChuNastaseNilssonPapageorgakis}. \footnote{The same result could be inferred from the idea of the fractional (anti)M2-branes as probe (anti)M5-branes wrapped around $S^3/Z_k$ ($S^1$ fibration over $CP^1$); see \cite{ABJ}.} Third, the bulk settings break the inversion (and so, the special conformal transformation $K_\mu$) symmetry and scale-invariance (denoted by the dilation operator $D$) because of the mass and nonlinear terms in the bulk action and translational-invariance (denoted by the translation operator $P_\mu$) because of non-constant solutions. \footnote{Meantime, although because of the breaking of scale-invariance, the boundary operators obtain anomalous dimensions (due to corrections to the bulk tree-level diagrams and presence of interactions), here we consider their bare dimensions in quenched approximation.} As a result, the conformal symmetry $SO(4,1)$ (as the isometry of $EAdS_4$) breaks into $SO(4)$, which in turn includes six generators consisting of three Lorentz transformations (denoted by the operator $L_{\mu \nu}$) and $R_\mu \approx \left(K_\mu + a^2 P_\mu \right)$ \footnote{It is noticeable that with the bulk solution (\ref{eqA20}), $S_\mu \approx \left(K_\mu - b_0^2\, P_\mu \right) $ is used instead.} corresponding to rotations on $S^3$, where $a$ is the scale parameter. The four generators of the broken symmetries (translations and scale transformations)-and so the four free parameters $a$ (or $b_0$) and $\vec{u}_0$- move the $SO(4)$-symmetric ($SO(3, 1)$ in Lorentzian signature) bubble around in the 4D bulk.

On the other hand, the mass spectrum of 11D SUGRA over $AdS_4 \times S^7/Z_k$ \footnote{For original works on the spectrum, see for example \cite{Heidenreich1982}, \cite{Englert1983}, \cite{Sezgin1983}, \cite{Biran}, \cite{Casher1984} among many others and \cite{Duff1986} as a comprehensive review until then with references therein and also \cite{Halyo1}, \cite{HokerPioline} as well as \cite{Bianchi2}, \cite{Forcella.Zaffaroni}, \cite{Bobev-Bomans2021} for newer looks.} includes three generations of scalars ($0_1^+, 0_2^+, 0_3^+$) and two generations of pseudoscalars ($0_1^-, 0_2^-$). In fact, the massless multiplet ($n=0$) includes a graviton ($\textbf{1}$), a gravitino ($\textbf{8}_s$), 28 spin-1 fields ($\textbf{28}$), 56 spin-$\frac{1}{2}$ fields ($\textbf{56}_{s}$), 35 scalars ($\textbf{35}_{v}$) of $0_1^+$ arising from the external components ($\mathcal{A}_{\mu \nu \rho}$) and 35 pseudoscalars ($\textbf{35}_{c}$) of $0_1^-$ arising from the internal components ($\mathcal{A}_{m n p}$), without any $H$-singlet under the branching $G\rightarrow H$ for scalars ($\textbf{35}_v \rightarrow \bar{\textbf{10}}_{-2} \oplus \textbf{10}_{2} \oplus \textbf{15}_{0}$) and pseudoscalars ($\textbf{35}_c \rightarrow \textbf{10}_{-2} \oplus \bar{\textbf{10}}_{2} \oplus \textbf{15}_{0}$).\footnote{It is noticeable that after the Hopf reduction ${S^7}/{Z_k} \rightarrow CP^3 \ltimes S^1/Z_k$, just the neutral states under $U(1)$ remain in the spectrum \cite{NilssonPope}, and the states with odd $n$ on $S^7$ are excluded.} In massive or higher KK multiplets ($n>0$), the massless ($m^2=0$) pseudoscalar and scalar set in $\acute{\textbf{840}}_{s}$ of $0_1^-$ with $n=2$ and $\textbf{1386}_{v}$ of $0_1^+$ with $n=4$ of $G$, again without any $H$-singlet under the branching. As the same way, the massive ($m^2=18$) pseudoscalar sets in $\acute{\textbf{840}}_{c}$ of $0^-_2$ with $n=4$ and $\textbf{75075}_{vc}$ of $0^-_1$ with $n=6$ of $G$, while as scalar it sets in $\textbf{30940}_{v}$ of $0^+_1$ with $n=10$ and  $\textbf{23400}_{v}$ of $0^+_3$ with $n=6$ and also $\textbf{1}$ of $0^+_2$ with $n=2$ of $G$, again without any $H$-singlet under the branching, except for the last one $ \textbf{1}(0,0,0,0) \rightarrow \textbf{1}_{0}[0, 0, 0]$.  

However, because of the triality of $G$ \footnote{Look at \cite{GustavssonRey} for related studies with the triality.}, one can exchange its three inequivalent reps $\textbf{8}_v$, $\textbf{8}_s$, $\textbf{8}_c$; In fact, to find the desired singlet modes and realize SUSY breaking in the boundary theory, we swap the three reps \footnote{It is reminded that $\textbf{8}_s \rightarrow \textbf{1}_{-2} \oplus \textbf{1}_{2} \oplus \textbf{6}_{0}$, $\textbf{8}_c \rightarrow \textbf{4}_{-1} \oplus \bar{\textbf{4}}_{1}$, $\textbf{8}_v \rightarrow \bar{\textbf{4}}_{-1} \oplus \textbf{4}_{1}$ under the branching $G \rightarrow H$.}. Therefore, with the swapping $\textbf{8}_s \leftrightarrow \textbf{8}_c$ and $\textbf{8}_v$ fixed, which means exchanging spinors(supercharges) with fermions and keeping scalars unchanged, the massless and massive pseudoscalar reps change accordingly without any $H$-singlet under the branching of the resulting reps, while the scalar reps do not change. As the same way, after the swapping $\textbf{8}_s \leftrightarrow \textbf{8}_v$ and $\textbf{8}_c$ fixed, which means exchanging spinors with scalars and keeping fermions unchanged, the resulting reps of both modes as pseudoscalar do not include any $H$-singlet under the branching. However, we have $\textbf{1386}_{s}$ and $\textbf{30940}_{s}$, $\textbf{23400}_{s}$ from the massless and massive scalar modes, respectively, while the rep $\textbf{1}$ of $0^+_2$ with $n=2$ of $G$ remains the same as before, with the latter swapping. For the latter reps, the branching $G \rightarrow H$ reads
\begin{equation}\label{eq48}
   \begin{split}
     & \ \ \ \ \ \ \ \ \ \ \ \ \ \ \ \ \ \ \ \ \ \  \ \ \textbf{30940}_{s} \rightarrow \textbf{1}_{0} \oplus \acute{\textbf{20}}_{0} \oplus \textbf{105}_{0} \oplus \textbf{336}_{0} \oplus \acute{\textbf{825}}_{0} \oplus \acute{\textbf{1716}}_{0} \oplus \textbf{3185}_{0} \oplus ... \, , \\
     & \ \ \ \ \ \ \ \ \ \ \ \ \ \ \ \ \ \ \ \ \ \  \ \ \textbf{23400}_{s} \rightarrow  \textbf{1}_{0} \oplus \textbf{15}_{0} \oplus 3\, (\acute{\textbf{20}}_{0}) \oplus \textbf{84}_{0} \oplus 3\, (\textbf{105}_{0}) \oplus 2\,({\textbf{175}}_{0}) \\
     & \ \ \ \ \ \ \ \ \ \ \ \ \ \ \ \ \ \ \ \ \ \  \ \ \ \ \ \ \ \ \ \ \ \ \ \ \ \ \ \ \ \, \oplus {\textbf{336}}_{0} \oplus {\textbf{729}}_{0} \oplus 2\,({\textbf{735}}_{0}) \oplus {\textbf{3640}}_{0} \oplus ...\ ,
   \end{split}   
\end{equation}
where we have only written $U(1)$-neutral reps, and that the corresponding reps for $\textbf{1386}_{s}$ are the same as the first four terms of the reps above for $\textbf{30940}_{s}$ under the branching. As a result, we see that after exchanging $\textbf{s} \leftrightarrow \textbf{v}$, there is the desired $H$-singlet rep ($\textbf{1}_{0} $) for both massless and massive (pseudo)scalars we consider here.

On the other hand, a bulk (pseudo)scalar with near the boundary behaviour of (\ref{eq32}) could be quantized with either the Neumann or alternate ($\delta\beta =0$) boundary condition for the masses in the range of $-9/4\leq m^2 \leq -5/4$ or the Dirichlet or standard ($\delta\alpha=0$) boundary condition that can in turn be applied to any mass (see for instance \cite{KlebanovWitten} and \cite{Balasubramanian02}), while the regularity (that $\Delta_+$ is real) and stability require that the mass is above the Breitenlohner–Freedman (BF) bound $m^2\geq m_{BF}^2=-9/4$ \cite{BreitenlohnerFreedman}, \cite{Duff1984-3}. As a result, for the massless and massive modes, only mode $\beta$ is normalizable; and $\alpha$ and $\beta$ have holographic expositions as source and vacuum expectation value of the one-point function of the operator $\Delta_+$, and vice versa for the operator $\Delta_-$. Then, we write the Euclidean AdS/CFT dictionary as
\begin{equation}\label{eq49}
   \begin{split}
   &  \langle \mathcal{{O}}_{\Delta_+} \rangle_{\alpha} = - \frac{\delta W[\alpha]}{\delta\alpha} = \beta, \quad \langle \mathcal{{O}}_{\Delta_-} \rangle_{\beta} = - \frac{\delta \tilde{W}[\beta]}{\delta\beta}= \alpha, \\
   & \ \ \ \ \ \ \ \ \ \ \ \ \ \ \tilde{W}[\beta] = - W[\alpha] - \int d^3 \vec{u}\ \alpha(\vec{u})\, \beta(\vec{u}),
   \end{split}
\end{equation}
where $W[\alpha]$ ($\tilde{W}[\beta]$) is the generating functional of the connected correlator of the operator $\mathcal{{O}}_{\Delta_+}$ ($\mathcal{{O}}_{\Delta_-}$) on the usual (dual) boundary CFT$_3$ with ${\Delta_+}$ ($\Delta_-$) quantization.

\section{Dual Solutions in Boundary 3D Field Theories} \label{section05}
The bulk setups with the symmetries  discussed in the previous section \ref{section04}, including parity breaking, are dual to the boundary CS $O(N)$ or $U(N)$ interacting vector models.\footnote{It is noticeable that according to \cite{Vasiliev2012}, nonlinear Higher-Spin gauge theories violating parity in $AdS_4$ correspond to nonlinear interacting 3D boundary CFTs.} However, we usually consider elements of ABJM's model with, depending on the case, only one scalar (say $Y =\varphi= h(r)\,\emph{\textbf{I}}_N$, with $h(r)$ as the scalar profile) or fermion (say $\psi$) \footnote{The singlet (pseudo)scalar or fermion we consider could be taken from decomposing the eight (pseudo)scalars or fermions as $X^I \rightarrow (\Phi^n, \Phi, \bar{\Phi})$, with $\Phi$ representing either $\psi$ or $Y$, $I,J...=(1,...6,7,8)=(n,7,8)$ and $\Phi=\Phi^7+i\, \Phi^8$, $\Phi^\dagger=\bar{\Phi}$, transforming in the rep $(\textbf{6}_{0}, \textbf{1}_{2}, \textbf{1}_{-2})$ under $SO(8)\rightarrow SU(4)_R \times U(1)_b$.} resulting in zero scalar and fermion potentials, and a deformation as
\begin{equation}\label{eq50}
  \mathcal{L}^{(p)} = \mathcal{L}_{CS}^+ - \texttt{tr} \left(i \bar{\psi}\, \gamma^k D_k \psi \right) - \texttt{tr}\left(D_k Y^{\dagger} D^k Y \right) - \mathcal{W}_\Delta^{(p)},
\end{equation}
where the CS Lagrangian reads
\begin{equation}\label{eq50a}
  \mathcal{L}_{CS}^+ = \frac{i k}{4\pi}\ \varepsilon^{ij k}\ \texttt{tr} \left(A_i^+ \partial_j A_k^+ + \frac{2i}{3} A_i^+ A_j^+ A_k^+ \right),
\end{equation} 
which is attributed to the remaining $U(1)$ part of the original quiver gauge group discussed in section \ref{section04} \footnote{We may also take the $U(1) \times U(1)$ part of the gauge group, with the gauge fields $A_i^{\pm} \equiv (A_i \pm \hat{A}_i)$, noting that the fundamental fields of ABJM are neutral wrt $A_i^+$ (diagonal $U(1)$) and $A_i^-$ acts as baryonic symmetry; and since our (pseudo)scalars are neutral, we set $A_i^-=0$. As a result, we will also examine the sum of $\mathcal{L}_{CS}$ (for $A_i$) and  $\hat{\mathcal{L}}_{CS}$ (for $\hat{A}_i$) instead of $\mathcal{L}_{CS}^+$ in the boundary analyzes.}, $D_k \Phi = \partial_k \Phi + i A_k\, \Phi - i \Phi\, \hat{A}_k$, $F_{ij}=\partial_i A_j - \partial_j A_i + i \left[A_i, A_j \right]$, and $\mathcal{W}_\Delta^{(p)}$, whose integral is $W$ in (\ref{eq49}), stands for (with $p$ marking) deformations we make with various $H$-singlet operators. 

\subsection{Marginal Deformations and Solutions For the Massless State}\label{sub05-01}
For the bulk solutions in subsections \ref{section02.01} and \ref{section02.02}, arising from taking the backreaction, which correspond to (exactly and irrelevant \footnote{See footnote \ref{fotnot19}.}) marginal operators, besides the $\Delta_+=3$ operators of $\mathcal{O}_{3}^{(a)} =\texttt{tr}(\varphi \bar{\varphi})^3$, $\mathcal{O}_{3}^{(b)} =\texttt{tr}(\varphi \bar{\varphi}) \texttt{tr}(\psi \bar{\psi})$, $\mathcal{O}_{3}^{(c)}=\texttt{tr}({A} \wedge {F})$ and $ \mathcal{O}_3^{(d)}=\texttt{tr}(\varphi \bar{\varphi})\, \varepsilon^{ij}\, F_{ij}^+$ already considered in \cite{Me3}, \cite{Me4},\cite{Me5}, \cite{Me7}, \cite{Me8}, \cite{Me9} and \cite{Me10}, here we include two new ones: 
\begin{equation}\label{eq51}
 \mathcal{O}_{3}^{(e)}= \texttt{tr}(\varphi \bar{\varphi})^2\, \varepsilon^{k ij} \varepsilon_{ij}\, A_k^+, \quad \mathcal{O}_{3}^{(f)}= \texttt{tr}(\psi \bar{\psi})\, \varepsilon^{k ij} \varepsilon_{ij}\, A_k^+.
\end{equation}
Next, we consider a deformation as
\begin{equation}\label{eq52}
\mathcal{W}_3^{(abf)}=\lambda_6\, \mathcal{O}_{3}^{(a)} + \hat{\lambda}_6\, \mathcal{O}_{3}^{(b)} + \check{\lambda}_6\, \mathcal{O}_{3}^{(f)},
\end{equation}
where the $\lambda$'s are coupling constants, and set $\alpha = 1$ for now. Then, if we take both CS terms $\mathcal{L}_{CS}+ \hat{\mathcal{L}}_{CS}$ instead of $\mathcal{L}_{CS}^+ $ in (\ref{eq50}), after some mathematical manipulations
on the resultant scalar $\bar{\varphi}=\varphi^{\dag}$, fermion $\bar{\psi}$ and gauge $A_k^+$ field equations, we get
\begin{equation}\label{eq53a}
 \partial_k \partial^k \varphi -3\, \lambda_6\, \varphi^5 =0,
\end{equation}
\begin{equation}\label{eq53b}
 i\, \bar{\psi} \gamma^k \partial_k \psi + 2\, \bar{\psi} \gamma^k \psi\, A_k^+ + \frac{i k}{4\pi}\, \varepsilon^{ijk} F_{ij}^+\, A_k^+ =0,
\end{equation}
where $Y=Y^{\dag}$ and $A_i^-=0$ are also set. After that, a closed solution for (\ref{eq53a}) reads
\begin{equation}\label{eq54}
   h = \left( \frac{1}{g_6} \right)^{1/4} \left( \frac{a}{a^2 + (\vec{u}-\vec{u}_0)^2} \right)^{1/2},
\end{equation} 
where $g_6 \equiv -\lambda_6$, while by employing the ansatz 
\begin{equation}\label{eq55}
 A_k^+ = \varepsilon_{k ij}\, \varepsilon^{ij} A^+(r),
\end{equation}
where $A^+(r)$ is a scalar function on the boundary, a solution for (\ref{eq53b}) reads
\begin{equation}\label{eq56}
      A^+ = \frac{3}{4} \left( \frac{a}{a^2 + (\vec{u}-\vec{u}_0)^2} \right),
\end{equation}
\begin{equation}\label{eq57}
     \psi= \tilde{a} \left( \frac{{a} + i (\vec{u} - \vec{u}_0). \vec{\gamma} }{\left[ {a}^2 + (\vec{u} - \vec{u}_0)^2 \right]^{\varsigma=3/2}} \right) \chi,
\end{equation} 
where $\vec{\gamma}=(\sigma_2, \sigma_1, \sigma_3)$ are the Euclidean gamma matrices and $\chi$ with $\chi^{\dag} \chi=1$ is a constant dimensionless spinor\footnote{See \cite{Akdeniz1979} for a similar ansatz.}. Finally, from computing the corresponding boundary action
\begin{equation}\label{eq58}
   {S}_{(3)}^{modi.} = - \int \left( \left(\partial_i \varphi \right)^2 -2\, \texttt{tr}(\bar{\psi} \gamma^3 \psi) A_3^+ + \lambda_6\, \varphi^6 + \hat{\lambda}_6\, \varphi^2\, \texttt{tr}(\psi \bar{\psi}) + 12\,  \check{\lambda}_6\, A^+\, \texttt{tr}(\psi \bar{\psi}) \right)
\end{equation}
based on the solutions (\ref{eq54}), (\ref{eq56}) and (\ref{eq57}), setting the couplings equal to 1 and $\tilde{a}= a = a^{\dag}$ for simplicity, we get  
\begin{equation}\label{eq58a}
   {S}_{(3)}^{modi.} = - 36 \int_0^\infty \frac{\pi\, a^3\, r^2}{\left( {a}^2+r^2 \right)^3}\, dr = - \frac{9}{4}{\pi^2},
\end{equation}
\footnote{Note that we could take $r=|\vec{u}-\vec{u}_0|$ (or $|x-x_0|$) with $\vec{u}_0$ (or $x_0$) as an arbitrary origin. \label{fotnot18}} which is finite, indicating an instanton with size ${a} \geq 0$ at the origin ($\vec{u}_0=0$) of a three-sphere with radius $r$ at infinity ($S_\infty^3$). 

As a result and a basic test of the correspondence, wrt (\ref{eq49}), we have
\begin{equation}\label{eq59}
     \langle \mathcal{O}_{3}^{(a,b,f)} \rangle_{{\alpha}} = a_1 \left( \frac{{a}}{{a}^2+(\vec{u} - \vec{u}_0)^2} \right)^3,
\end{equation}
with $a_1, a_2,...$ as boundary constants, which is compatible with near the boundary behavior of (\ref{eq41}) with $\Delta_+ =3$, wrt (\ref{eq32}), in the limit of ${a}\rightarrow 0, r\rightarrow \infty$. Meantime, comparing with the bulk closed solution of (\ref{eq22}), this boundary solution may be considered as an instanton sitting at the conformal point of $u =a$. \footnote{In fact, we already discussed in \cite{Me8} and \cite{Me9} the potential $- \lambda_6 (\varphi^2)^3$ of the tri-critical $O(N)$ or $U(N)$ model that is unbounded from below and so, there are instabilities near the potential extrema and tunneling mediated by Fubini-like instantons of the size $a$ and locations $\vec{u}_0$. Meantime, for any positive value of $\lambda_6$, the corresponding operator is not exactly marginal but it becomes quantum irrelevant; see \cite{Pisarski1982}, \cite{Bardeen1984}, \cite{Papadimitriou2006} and \cite{CrapsHertog}. On the other hand, we saw in subsection \ref{section02.01} when taking the backreaction of the whole 11D space, the resultant bulk solutions corresponded to marginally irrelevant deformations. In other words, by quantum corrections and the breaking of conformal invariance, an exactly marginal configuration may change to a marginally irrelevant one. \label{fotnot19}} 

Note also that with $\hat{\lambda}_6=\check{\lambda}_6=0$ in (\ref{eq52}) and including a mass-deformation term ($m_b^2\, \texttt{tr}(\varphi \bar{\varphi})$), \footnote{For studies on massive deformations of ABJM model, see \cite{Gomiz}, \cite{Arav2019}, \cite{Kim2019}.} we have in general the RB model (look also at \cite{Me9}), whose $\bar{\varphi}$ equation reads
\begin{equation}\label{eq60}
\left( \partial_i \partial^i -m_b^2 \right) h + 3\, g_6\, h^5 = 0.
\end{equation}  
Solutions for its free massive equation are available in terms of (modified) Bessel functions, with an explicit one as
\begin{equation}\label{eq61}
h_c(r) \cong \frac{a_2}{\sqrt{m_b}} \frac{e^{-m_b\, r}}{r},
\end{equation}
which satisfies the condition $h_c(r\rightarrow \infty)\rightarrow 0$, resulting in a finite action. Solutions for the interaction equation could be obtained in the context of constrained instantons; see \cite{Affleck1}, \cite{Nielsen1999}, \cite{KubyshinTinyakov01}
and \cite{GlerfossNielsen2005}. In fact, making use of (\ref{eq61}) as the initial data, one may employ perturbative methods and get solutions with a simple structure like $h \sim 1/r$ and so, we have the single-operator correspondence $\langle \mathcal{O}_{3}^{(a)} \rangle_{\alpha} \sim {1}/{r^6}$ with the typical near the boundary solution of (\ref{eq41}) for $\Delta_+ =3$. 

In particular, if we use just the operator $\mathcal{O}_{3}^{(e)}$ to deform the action of (\ref{eq50}) with, discarding its fermion kinetic term, the equations for $\bar{\varphi}$ and $A_k^+$ read
\begin{equation}\label{eq62a}
 \partial_i \partial^i \varphi -2 \varphi\, \texttt{tr}(\varphi \bar{\varphi})\, \varepsilon^{k ij} \varepsilon_{ij}\, A_k^+=0,
\end{equation}
\begin{equation}\label{eq62b}
 \frac{i k}{4\pi} \varepsilon^{k ij} F^+_{ij} - \texttt{tr}(\varphi \bar{\varphi})\, \varepsilon^{k ij} \varepsilon_{ij} + i \left[\varphi  \left(\partial^k \bar{\varphi} \right)- \left(\partial^k \varphi \right) \bar{\varphi}\right]=0,
\end{equation}
respectively. Next, with $\varphi=\bar{\varphi}$, from the last two equations, we can write
\begin{equation}\label{eq63}
 \partial_i \partial^i h(r)=0 \Rightarrow h(r)= {a}_3 + \frac{{a}_4}{r},
\end{equation}
while for the gauge field, we may use the ansatz (\ref{eq55}) with $A^+(r) \sim 1/r^2$ and so, with ${a}_3=0$, the basic correspondence  $\langle \mathcal{O}_{3}^{(e)} \rangle_{\alpha}\sim {1}/{r^6}$ is realized, with near the boundary solution of (\ref{eq41}) with $\Delta_+ =3$, wrt (\ref{eq32}). On the other hand, if $\varphi \neq \bar{\varphi}$, which is allowed due to being in Euclidean space, and explicitly with 
\begin{equation}\label{eq64}
 \varphi = h(r)\, I_N, \quad \varphi^{\dag} = {a}_5\, I_N,
\end{equation}
from the equations (\ref{eq62a}) and (\ref{eq62b}), we get
\begin{equation}\label{eq65}
 \partial_k \left(\varepsilon^{k ij} F^+_{ij} \right)=0, \quad  A_k^+ =  i\, \partial_k \ln h.
\end{equation}
The latter solution is reminiscent of the duality (${A}_k=\eta_{k j}\, \partial^j h/h$) between the instanton solution of the pure $SU(2)$ Yang-Mills theory \cite{Belavin},
\begin{equation}\label{eq66}
{A}_k \approx \eta_{k j}\, \frac{(x-x_0)^j}{a^2+(x-x_0)^2} \Rightarrow {F}_{i j} \approx \eta_{i j} \left(\frac{a}{a^2+(\vec{u}-\vec{u}_0)^2}\right)^2,
\end{equation}
with $\eta_{i j}$ as 't Hooft symbols \cite{Hooft1976}, and the $SO(4)$-invariant solution of the so-called $\varphi^4$ model as \footnote{See for instance \cite{Fubini1}, \cite{Lipatov1977}, \cite{brzein1977}, \cite{McKane-Wallace1978}, \cite{Actor1982}, \cite{Alfaro1976} and \cite{Corrigan-Fairlie1977}.} 
\begin{equation}\label{eq67}
 \nabla^2 h + \lambda_4\, h^3=0 \Rightarrow 
 h = \sqrt{\frac{8}{\lambda_4}} \left( \frac{a}{a^2 + (\vec{u}-\vec{u}_0)^2} \right).
\end{equation}
As a result, with the latter solutions, we have the same correspondence as (\ref{eq59}) for $\mathcal{O}_{3}^{(e)}$. 

\subsection{Irrelevant Deformations and Solutions For the Massive State}\label{sub05-01}
For the Higgs-like (pseudo)scalar $m^2=18$, besides the $\Delta_+=6$ operators introduced in \cite{Me7}, including $\mathcal{O}_{6}^{(b)}= (\mathcal{O}_{3}^{(b)})^2$ (a double-trace deformation), we consider a few new ones. The operator $\mathcal{O}_{6}^{(a)}=\texttt{tr}(\psi \bar{\psi})^3$, first taken in \cite{Me9}, is interesting in that it can also be taken in the CF model. In fact, if we take a deformation as 
\begin{equation}\label{eq68}
\mathcal{W}_6^{(a)}=m_f\, \mathcal{O}_{2}^{(a)} + \tilde{g}_6\, \alpha\, \mathcal{O}_{6}^{(a)},
\end{equation}
where $\mathcal{O}_{2}^{(a)}=\texttt{tr}(\psi \bar{\psi})$, by excluding the scalar kinetic term in (\ref{eq50}), the $\bar{\psi}$ equation reads
\begin{equation}\label{eq69}
 i\, \gamma^k \partial_k \psi + m_f\, \psi +3\, \tilde{g}_6\, \alpha\, \psi\, \texttt{tr}(\psi \bar{\psi})^2 =0. 
\end{equation} 
The solution of (\ref{eq57}) with $\tilde{a}=a/(\tilde{g}_6)^{1/4}$ is also valid for the latter equation provided that
\begin{equation}\label{eq69a}
    \alpha = \texttt{tr}(\psi \bar{\psi})^{-3/2} 
\end{equation} 
for the massless ($m_f = 0$) and with $m_f \rightarrow \tilde{\alpha}(\vec{u})= \texttt{tr}(\psi\bar{\psi})^{1/2}$ for the massive case; and so, the deformation might in fact be considered as a triple-trace one. As a result,
\begin{equation}\label{eq70}
     \langle \mathcal{O}_{6}^{(a)} \rangle_{{\alpha}} = \left( \frac{\tilde{a}}{{a}^2+(\vec{u} - \vec{u}_0)^2} \right)^6,
\end{equation}
which, wrt (\ref{eq32}), corresponds to (\ref{eq47}) for $\Delta_+=6$ with (\ref{eq47a}) and $a_0^2-b_0^2=a^2$, $- a_0\, b_0 \sim \tilde{a}^2$. Moreover, we can obtain an explicit profile for $f(r)$ in (\ref{eq37}) from this solution. In fact, according to the above discussions, with $\Delta_\pm =6, -3$ and the correspondence rules of (\ref{eq49}), we get
\begin{equation}\label{eq71}
   f(r)= \left[ \frac{\tilde{a}}{a^2 + r^2} \right]^{1/2},
\end{equation}
where $C_- =\sqrt{C_+}$ and set $C_+=1$ for simplicity.

It is also interesting to check the BF duality (or 3D Bosonization)- see for instance \cite{Choudhury2018} and \cite{Aharony2018}- from our setups attributed to RB and CF models \footnote{See \cite{Me9}, where we have used these models in more detail.} at the level of the solutions. Indeed, under the BF duality, the coupling of $\tilde{g}_6\, \texttt{tr}(\psi \bar{\psi})^3\sim \tilde{g}_6\, \sigma_f^3$, where $\sigma_f$ is the so-called Hubbard-Stratonovich field, is mapped into the coupling of ${g}_6\, \texttt{tr}(\varphi \bar{\varphi})^3 $ ($\mathcal{W}_3^{(a)}$ of (\ref{eq52})); see \cite{Minwalla-Yokoyama2015} and \cite{Giombi2017}.\footnote{Indeed, a double-trace deformation of the latter takes the RB model to the CF model; see the deformation (\ref{eq68}).} In this regard, from the solutions (\ref{eq54}) of the boson model and (\ref{eq57}) of the fermion model, we have 
\begin{equation}\label{eq72}
     \texttt{tr}({\psi} \bar{{\psi}}) = \left(\frac{1}{\tilde{g}_6} \right)^{3} \left( \frac{a}{{a}^2+(\vec{u} - \vec{u}_0)^2} \right)^6= \texttt{tr}(\varphi \bar{\varphi})^2,
\end{equation}
where $\tilde{g}_6 \leftrightarrow g_6$ and so, realizing the BF duality with $\psi \leftrightarrow \varphi^2$, or with $\psi \leftrightarrow \varphi$ when including $\alpha$ in the fermion model.

In this way, we now examine two new operators:
\begin{equation}\label{eq73}
 \mathcal{O}_{6}^{(c)}= \texttt{tr}(\psi \bar{\psi})^2\, \varepsilon^{ij}\, F_{ij}^+, \quad \mathcal{O}_{6}^{(d)}= \texttt{tr}(\psi \bar{\psi})\, \texttt{tr}(F_{ij}^+\, F^{+\, ij});
\end{equation}
with the associated deformations
\begin{equation}\label{eq74}
\mathcal{W}_6^{(q)}=\alpha\, \mathcal{O}_{6}^{(q)},
\end{equation}
where $q=c,d,...,h$ from now on. Next, discarding the scalar kinetic term of (\ref{eq50}), with $\mathcal{W}_6^{(c)}$, the fermion $\bar{\psi}$ and gauge $A_i^+$ equations read
\begin{equation}\label{eq75}
 i\, \gamma^k \partial_k \psi + 2\, \alpha\, \psi\, \texttt{tr}(\psi \bar{\psi}) \varepsilon^{ij} F_{ij}^+ =0,
\end{equation}
\begin{equation}\label{eq76}
 \frac{i k}{4\pi} \varepsilon^{k ij} F^+_{ij} + 2\, \bar{\psi}\, \gamma^k\, \psi= 0,
\end{equation}
respectively, reminding that the second term on the LHS of (\ref{eq76}) exists when we include both CS terms (for $A_i$ and $\hat{A}_i$) in (\ref{eq50}) and that $F_{ij}^-=0, A_i^-=0$ is set.  With just the CS term of (\ref{eq50a}), the ansatz
\begin{equation}\label{eq77}
      A_\mu^+ = \omega_{\mu \nu}\, x^\nu A(r), \quad \omega_{\mu \nu}=
  \left\{ \begin{split}
  & 1 \ \ \ : \ \nu>\mu, \\
  & 0 \ \ \ : \ \nu=\mu,\ \ \mu,\nu \neq i, j ,
  \end{split} \right.
\end{equation} 
for the $U(1)$ gauge field, with $\mu,\nu$ for the boundary indices as well and $A(r)$ as another boundary scalar function, gives us a desired solution- see also \cite{Me10}- as
\begin{equation}\label{eq78}
      A(r)= \frac{{a}_6 + 4\, a_7\, r}{4\, r^4} \Rightarrow \varepsilon^{ij} F_{ij}^+ \equiv F^+ = \frac{a_6}{r^4}.
\end{equation} 
In this case, a solution for $\psi$ is read out of (\ref{eq57}) with $a=0$ and $\tilde{a} = \frac{i}{2} \sqrt[3]{\frac{4}{5}}$; and as a result, 
\begin{equation}\label{eq79}
     \langle \mathcal{O}_{6}^{(c)} \rangle_{\alpha} = \frac{a_6\, \tilde{a}^4}{r^{12}} \cong f(r), 
\end{equation}
with $f(r)$ in (\ref{eq44}); and one can also adjust $\bar{C}_6 =a_6\, \tilde{a}^4$ of (\ref{eq45}), wrt (\ref{eq32}).

However, from the combination of (\ref{eq75}) and (\ref{eq76}), we get
\begin{equation}\label{eq80}
 \gamma^k\, \partial_k \psi + \frac{16 \pi}{k}\, \alpha\, \texttt{tr}(\psi \bar{\psi})^2\, \gamma^3\, \psi =0,
\end{equation}
where taking the third component of the gamma matrices is for compatibility with the solution we take for $\psi$, which in turn reads from (\ref{eq57}) with $\tilde{a} =a \sqrt[1/2]{\frac{-3 i k}{16 \pi}}$ and so, from (\ref{eq75}), we have
\begin{equation}\label{eq81}
      F_{ij}^+ = a_8\, \varepsilon_{ij} \left( \frac{a}{{a}^2+(\vec{u} - \vec{u}_0)^2} \right)^2,
\end{equation} 
where $a_8=(3 \pi i/k)^{1/2}$, reminding that $F^+(r\rightarrow \infty) \rightarrow 0$.\footnote{It is noticable that the $A_i^+$ equation of (\ref{eq76}) and the solution of (\ref{eq57}) result in zero magnetic charge or flux, $\Phi=\oint_{S^3_\infty} F^+=0$; see also \cite{Me3}.} As a result,
\begin{equation}\label{eq82}
     \langle \mathcal{O}_{6}^{(c)} \rangle_{{\alpha}} = \frac{3}{2} \frac{a^2\, \tilde{a}^2}{\left[a^2+(\vec{u} - \vec{u}_0)^2\right]^6},
\end{equation}
which, for $\Delta_+=6$, can be made to correspond to (\ref{eq47}) with (\ref{eq47a}) and also to (\ref{eq22}) with an instanton at the conformal point of $u=a$.

Moreover, to confirm the instanton nature of the Euclidean solutions, we compute the value of the corresponding action as
\begin{equation}\label{eq83}
     S_{(6c)}= \int \mathcal{W}_6^{(c)}\, d^3\vec{u} \quad \Rightarrow S_{(6c)}^{modi.}= \frac{3 \pi^2}{8} \frac{a}{\tilde{a}}, 
\end{equation} 
where we have used the result of the integral in (\ref{eq58a}) and the same interpretation. 

Similarly, for the deformation $\mathcal{W}_6^{(d)}$ of (\ref{eq74}), discarding the scalar kinetic term of (\ref{eq50}) and taking both CS terms, the fermion $\bar{\psi}$ and gauge $A_i^+$ equations read 
\begin{equation}\label{eq84}
 i\, \gamma^k \partial_k \psi +  \alpha\, \psi\, \texttt{tr}(F_{ij}^+\, F^{+\, ij}) =0,
\end{equation}
\begin{equation}\label{eq85}
 \frac{i k}{4\pi} \varepsilon^{k ij} F^+_{ij} + 2\, \bar{\psi}\, \gamma^k\, \psi +4\, \alpha\, \texttt{tr}(\psi \bar{\psi})\, \partial_j F^{+\, jk}= 0.
\end{equation}
Then, using (\ref{eq69a}), solutions for the fermion and gauge fields are read from (\ref{eq57}) with $\tilde{a} =a \sqrt[1/2]{\frac{-9 i k}{8 \pi}}$ and from (\ref{eq81}) with $a_8=1$, respectively. However, if we set $\alpha=1$ in the equations, a solution for $\psi$ is read from (\ref{eq57}) with $\varsigma=0$ instead of $3/2$ along with the gauge solution (\ref{eq81}) with $a_8=1/a$ to have $\tilde{a}$ the same as before. As a result, we have 
\begin{equation}\label{eq86}
     \langle \mathcal{O}_{6}^{(d)} \rangle_{{\alpha}} =  \frac{a^2\, \tilde{a}^2}{\left[a^2+(\vec{u} - \vec{u}_0)^2\right]^3},
\end{equation}
which corresponds to the bulk near the boundary solution of (\ref{eq42}) with $\check{C}_{-3}=0$ and $\hat{C}_{6}=a^2\, \tilde{a}^2$, and of course in the limit of ${a}\rightarrow 0, r\rightarrow \infty$, reminding the footnote \ref{fotnot18}.  

Another operator we consider is 
\begin{equation}\label{eq87}
 \mathcal{O}_{6}^{(e)}= \texttt{tr}(\varphi \bar{\varphi})^4\, \varepsilon^{ij}\, F_{ij}^+;
\end{equation} 
and deform the action of (\ref{eq50}), discarding its fermion term, with (\ref{eq74}) with $q=e$. As a result, the scalar $\bar{\varphi}$ equation reads
\begin{equation}\label{eq88}
 \partial_i \partial^i \varphi - 4\, \alpha\, \varphi\, \texttt{tr}(\varphi \bar{\varphi})^3\, \varepsilon^{ij}\, F_{ij}^+=0, 
\end{equation}
and the gauge $A_i^+$ equation is the same as (\ref{eq62b}) except leaving out the middle term on the LHS of it. Next, with $\varphi = \bar{\varphi}$, we can take for the gauge part a similar solution to (\ref{eq81}) with $a_8=1$ and then, taking $\alpha \sim \texttt{tr}(\varphi \bar{\varphi})^{-3}$ and $F^+ \sim h^4$, we obtain a similar solution to (\ref{eq54}) for $h$  with $g_6=1$ and so, the same correspondence as (\ref{eq70}) with $a=\tilde{a}$ for $\mathcal{O}_{6}^{(e)}$ is confirmed.

However, when $\varphi \neq \bar{\varphi}$ we take (\ref{eq64}) and next, from (\ref{eq88}) and (\ref{eq62b}) without the middle term on the LHS of the latter, we can write
\begin{equation}\label{eq89}
 \partial_k \left(\varepsilon^{k ij} F^+_{ij} \right) - \frac{16 \pi}{k}\, \alpha\, \texttt{tr}(\varphi \bar{\varphi})^4\, \varepsilon^{ij} F^+_{ij}=0.
\end{equation}
Then, using the ansatz (\ref{eq77}) and so $=F^+= -2 \left(6 A(r) + 2\, r \acute{A}(r) \right)$, with $a_5 =\sqrt[1/2]{\frac{-3 \sqrt{3} k}{16 \pi}}$ and $\alpha=1$, we get 
\begin{equation}\label{eq90}
  \frac{d^2 A(r)}{dr^2} + \left( \tilde{h}(r) + \frac{4}{r} \right) \frac{dA(r)}{dr} + \frac{3}{r}\, \tilde{h}(r)\, A(r)=0,
\end{equation}
for which we can write a solution as   
\begin{equation}\label{eq91}
 h^4(r) \equiv \tilde{h}(r) = \frac{n}{r} \Rightarrow A(r) = \frac{{a}_7}{r^3} + \frac{{a}_9}{r^n} \Rightarrow F^+ =\frac{4\, a_9}{r^n} (n-3),
\end{equation}
with $n$ as a real number. As a result, we have 
\begin{equation}\label{eq92}
     \langle \mathcal{O}_{6}^{(e)} \rangle_{{\alpha}} = \frac{4\,n\, a_5^4\, a_9}{r^{n+1}} (n-3),
\end{equation} 
which with $n=5$, wrt (\ref{eq32}), corresponds to the normalizable part of the bulk solution (\ref{eq42}), with adjusting the constants of both sides. As the same way, with $n=11$, it can be made to correspond to $f(r)$ in (\ref{eq44}).

Among other similar operators, if we use any of the following three operators
\begin{equation}\label{eq73}
 \begin{split}
 \mathcal{O}_{6}^{(f)} = \texttt{tr}(\varphi \bar{\varphi})\, & \texttt{tr}(\psi \bar{\psi})^2\, \varepsilon^{ijk}\, \varepsilon_{ij}\, A_k^+,  \quad \mathcal{O}_{6}^{(g)} = \texttt{tr}(\varphi \bar{\varphi})^2\, \texttt{tr}(\psi \bar{\psi})\, \varepsilon^{ij}\, F_{ij}^+,  \\  & \mathcal{O}_{6}^{(h)} = \texttt{tr}(\varphi \bar{\varphi})\, \texttt{tr}(\psi \bar{\psi})\, \varepsilon^{ijk}\, F_{ij}^+\, A_k^+, 
 \end{split}
\end{equation}
to deform (\ref{eq50}) with, wrt (\ref{eq74}), the fermion solution may be (\ref{eq57}) with $a=0$, the scalar solution may be (\ref{eq63}) and the gauge solution may be $F^+(r)$ in (\ref{eq78}) and $A^+(r) \sim 1/r^2$ (for $\mathcal{O}_{6}^{(f)}$) according to the ansatz of (\ref{eq55}). As a result and a primary test of the correspondence, $\langle \mathcal{O}_{6}^{(f,g,h)} \rangle_{{\alpha}} \sim 1/r^{12}$ matches with the bulk solution (\ref{eq44}) and (\ref{eq45}) for $\Delta_+=6$ as before.

\section{Summary and Comments}
In this article, we started from 11D SUGRA with the background geometry $AdS_4 \times S^7/Z_k$ fixed and a dynamical 4-form ansatz, and got a consistent truncation in that the resulting scalar equations in the external $EAdS_4$ space do not include any dependence to the internal space ingredients and the associated (pseudo)scalars are $H$-singlets. In addition, as the solutions are owned to probe (anti)M-branes wrapped around the three internal directions $CP^1 \ltimes S^1/Z_k$ in the (WR)SW background, they break all SUSYs and parity and the resultant theory is for anti-M2-branes. The scale-invariance is also broken due to mass terms and nonlinearities of the equations. Taking the backreaction, we got the massless ($m^2=0$) and massive ($m^2=1/9, 2/9$) modes corresponding to the exactly marginal and marginally irrelevant operators on the 3D boundary and then, wrote a closed solution for the resultant equation and computed its correction to the bulk background action. As well as, for the NPDE equation of the Higgs-like ($m^2=18$) mode, arisen from spontaneous symmetry breaking, we employed the ADM method and arrived at interesting series solutions appropriate for near the boundary analyzes. 

In order to realize the supersymmetry and parity breaking and also the realization of $H$-singlet bulk (pseudo)scalars, we swapped the three fundamental reps of $SO(8)$ and saw that under the branching of $G\rightarrow H$ such (pseudo)scalars are realized. Because of the bulk symmetries, the boundary duals could be come off in the singlet sectors of ABJM-like models, from which we built some new marginal and irrelevant operators composed of a scalar, a fermion and a $U(1)$ gauge field. Having said that, we saw that solutions with finite actions and $SO(4)$ symmetry on a three-sphere at infinity could be found. After that, we confirmed the state-operator correspondence, adjusted the bulk and boundary parameters and also specified unknown functions in the bulk from the boundary solutions. In addition, we confirmed a type of BF duality ($\varphi \leftrightarrow \psi$) between RB and CF models in terms of the solutions and correspondence. 

In order to further confirm the results and reconcile with previous studies by others and their applications, a few more points are worth mentioning. First, we remind that the instantons here are mainly attributed to the unbounded boundary potential from below and have also dual interpretations in the form of Coleman-de Luccia (CdL) bounces \cite{Coleman1980a} mediating false-vacuum decay and the formation of true-vacuum bubble within it; and according to \cite{Abbott1985}, such $AdS_4$ bubbles collapse and eventually end in a big crunch singularity.\footnote{We remind that CFTs with unbounded potentials from below have observables that evolve to infinity in finite time, and their bulk duals are gravities coupled to (pseudo)scalars with potentials coming from consistent truncations of supergravity, as it stands here; see for instance \cite{Hertog-Horowitz2005}, \cite{Craps2012} and \cite{BarbonRabinovici011}.} Second, that our bulk solutions (not) considering the backreaction correspond to (irrelevant) marginally irrelevant deformations (see also the footnote \ref{fotnot19}) is consistent with the result in \cite{Maldacena010} that says field theories on $dS_3$ with $SO(3,1)$-invariant solutions and irrelevant deformations are dual to vacuum decays and cosmic singularities in $AdS_4$. Third, we notice the probe (anti)M2-branes wrapped around $S^3/Z_k$ that result in domain-walls interpolating among different vacua \cite{Bena}; and according to \cite{Garriga2010}, a domain-wall at $u=0$ separates two degenerate $AdS_4$ vacua. Indeed, with conformal-invariance breaking, we deal with the problem on constant-$u$ patches, where the boundary is $dS_3$ in Lorentzian signature. Fourth, such a truncation is interesting in some cosmological (inflationary and bouncing) models \footnote{See, for instance, \cite{BumHoonLee2006} with references therein.}; In fact, our almost degenerate double-well scalar potential from (\ref{eq08}) accepts \emph{bounce} solutions and so, it is possible to address the problem from that point of view and provide interesting analyzes.

\begin{appendices}

\section{\large{EM Tensors and Resulting Equations}} \label{Appendix.A1}
we use the Einstein's equations 
\begin{equation}\label{eqA1} 
    \mathcal{R}_{MN} - \frac{1}{2} g_{MN} \mathcal{R} = 8 \pi\, \mathcal{G}_{11} T_{MN}^{{G}_4},
\end{equation}
where
\begin{equation}\label{eqA2}
 T_{MN}^{{G}_4} = \frac{1}{4!} \left[4\, {G}_{MPQR}\, {G}_N^{PQR} - \frac{1}{2} g_{MN}\, {G}_{PQRS}\, {G}^{PQRS} \right],
\end{equation} 
and the capital $M, N,...$, small $m, n,...$ and Greek $\mu, \nu,....$  indices are for the whole 11D, 6D internal $CP^3$ and 4D external $AdS_4$ spaces, respectively. 

Next, using the conventions and performing computations similar to (Appendix B of) \cite{Me8}, we get
\begin{equation}\label{eqA3}
 {G}_{PQRS}\, {G}^{PQRS} = 96 \left[ \frac{8}{3 R^8} \bar{f}_1^2  + \frac{R^2}{32} (\partial_\mu f_2)(\partial^\mu f_2) + \frac{1}{8} {f}_3^2\right],
\end{equation}
\begin{equation}\label{eqA4}
{G}_{\mu PQR}\, {G}_\nu^{PQR} = \frac{64}{R^8} \bar{f}_1^2\, g_{\mu \nu} + \frac{3 R^2}{4} (\partial_\mu f_2)(\partial_\nu f_2),
\end{equation}
\begin{equation}\label{eqA5}
{G}_{m PQR}\, {G}_n^{PQR} = \left[2 {f}_3^2 + \frac{R^2}{4} (\partial_\mu f_2)(\partial^\mu f_2) \right] g_{m n} ,
\end{equation}
\begin{equation}\label{eqA6}
{G}_{7 PQR}\, {G}_7^{PQR} = \frac{3 R^2}{4} (\partial_\mu f_2)(\partial^\mu f_2)\ g_{7 7},
\end{equation}
with a $4!$ factor for all terms. 

Then, by plugging (\ref{eqA3}) with (\ref{eqA4}), (\ref{eqA5}) and (\ref{eqA6}) back into (\ref{eqA2}), using (\ref{eq03}) with the conventions of (\ref{eq08a}), taking traces and using the Euler-Lagrange equation, we finally get
\begin{equation}\label{eqA13}
     \Box_4 f_2 + 4\, m^2\, f_2 \pm\, {4}\, \delta\, f_2^2 + 4\, \lambda\, f_2^3 \pm 4\, F=0,
\end{equation}
\begin{equation}\label{eqA14}
     \Box_4 f_2 + \left( 3\, m^2 - \frac{8}{R^2} \right) f_2 \pm\, 3\, \delta\, f_2^2 + 3\, \lambda\, f_2^3 \pm 3\, F=\pm\, \frac{8\, C_2}{R^3},
\end{equation}
\begin{equation}\label{eqA14a}
     \Box_4 f_2 - m^2\, f_2 \mp \delta\, f_2^2 - \lambda\, f_2^3 \mp F=0,
\end{equation}
for the external $AdS_4$ space, the internal $CP^3$ and eleventh $S^1/Z_k$ components respectively, noting that (\ref{eqA14a}) is the same as the main equation (\ref{eq08}).

\section{\large{Basics of ADM to Solve the Scalar Equation}} \label{Appendix.A2}
Adomian decomposition method or the inverse operator method \cite{Adomian1994} is a mathematical method especially for solving NPDEs; see also \cite{Wazwaz2009}. Because here we are looking for solutions near the boundary ($u=0$), we use 
\begin{equation}\label{eqA15}
f_0(0,r)= f(0,r) - u\, f_u (0,r), \quad f(0,r)= {f}(r)\, u^{\Delta_+}
\end{equation}
as initial data for the second order NPDE (\ref{eq29}), corresponding to Dirichlet boundary condition. In this way, from the main equation,  we can write \footnote{See also \cite{Me10}.}
\begin{equation}\label{eqA16}
     \Box_4 f_0 - m^2\, f_0 =0,
\end{equation}
with the closed solution of (\ref{eq22}), and
\begin{equation}\label{eqA17}
    \Box_4 f_{n+1} - m^2\, f_{n+1} = \sum_{n=0}^\infty A_n,
\end{equation}
where the nonlinear terms are written as the sum of Adomian polynomials $A_n$'s,
\begin{equation}\label{eq17a}
     A_n=\frac{1}{n!} \frac{d^n}{d\lambda^n} \left[\mathcal{F}\left(\sum_{n=0}^n \lambda^n\, f_n \right) \right]_{\lambda=0}, \quad n=0,1,2,... \ ,
\end{equation}
with $\mathcal{F}(f)$ for the nonlinear terms of (\ref{eq35}); and as a result,
\begin{equation}\label{eqA18}
    \begin{split}
    & \ A_0=6 f_0^3 - \delta\, f_0^2, \quad A_1= 18\, f_0^2\, f_1- 2\, \delta\, f_0^2\, f_1, \\
    & \quad A_2=18\, f_0^2\, f_2 + 18\,f_0\, f_1^2 - 2\, \delta\, f_0\, f_2- \delta\, f_1^2 , ... \ ,
     \end{split}
\end{equation}
where $\delta= 3\sqrt{3}\, m$. In this way, a series solution up to the $n$th order of the iteration processes may be written according to (\ref{eq36}).

On the other hand, with $f=(u/R_{AdS})\, g$ and $R_{AdS}=1$, from (\ref{eq29}), we can write 
\begin{equation}\label{eqeqA19}
\left(\partial_i \partial_i + \partial_u \partial_u \right) g_0 - 6\, g_0^3 =0,
\end{equation}
with the exact solution of
\begin{equation}\label{eqA20}
g_0(u,\vec{u}) ={\frac{2}{\sqrt{3}}}\, \left(\frac{b_0}{-b_0^2 + (u+a_0)^2 + (\vec{u}-\vec{u}_0)^2} \right),
\end{equation} 
which is indeed for the so-called conformally coupled (pseudo)scalar $m^2=-2$ in the SW version of (\ref{eq04}) with $C_3=1$- with $a_0, b_0$ as physically meaningful constants and remembering $|\vec{u}-\vec{u}_0| \equiv r$ when using the spherical coordinates- and
\begin{equation}\label{eqA21}
      \left(\partial_i \partial_i + \partial_u \partial_u \right)\, g_{n+1}- 6\, g_{n+1}^3 =\sum_{n=0}^\infty A_n,
\end{equation}
with the Adomian polynomials 
\begin{equation}\label{eqA21a}
     A_0 = \frac{(2+m^2)}{u^2}\, g_0 - \frac{3 \sqrt{3}\, m}{u}\, g_0^2, \quad A_1 = \frac{(2+m^2)}{u^2}\, g_1 - \frac{6 \sqrt{3}\, m}{u}\, g_0\, g_1, .... \ ,
\end{equation}
to get series solutions near the boundary $u=0$.

\end{appendices}

\end{document}